\documentclass[final,1p,12pt]{elsarticle}

\usepackage{amsmath,epsfig,latexsym}
\usepackage{axodraw}

\def\DRbar{{$\overline{DR}$}}
\def\ghat{{\hat{g}}}
\def\gtilde{{\tilde{g}}}

\def\MSbar{{$\overline{\mbox{MS}}$}}
\def\DRbar{{$\overline{\mbox{DR}}$}}

\def\modelname{{\tt MSSMdreg2dred}}
\def\model{{\tt \modelname.mod}}

\newcommand{\pleft}{P_{L}}
\newcommand{\pright}{P_{R}}
\newcommand{\chargec}{{\mathrm{C}}}

\newcommand{\Fermion}[1]{{\Psi_{#1}}}
\newcommand{\barFermion}[1]{{\overline{\Psi}_{#1}}}
\newcommand{\CCFermion}[1]{{\Psi_{#1}^{\chargec}}}
\newcommand{\barCCFermion}[1]{\overline{\Psi_{#1}^{\chargec}}}

\newcommand{\Gaugino}[1]{\tilde{\lambda}_{#1}}
\newcommand{\barGaugino}[1]{\bar{\tilde{\lambda}}_{#1}}
\newcommand{\gaugino}[1]{\lambda_{#1}}

\newcommand{\Vecbos}[3]{{V_{#1}^{#2}}_{#3}}
\newcommand{\scalar}[1]{{\phi_{#1}}}
\newcommand{\antiscalar}[1]{{\phi_{#1}^{\ast}}}

\newcommand{\restLL}[1]{\Lagr^{\text{trans}}_{#1}}

\newcommand*\clap[1]{\hbox to 0pt{\hss#1\hss}}

\newcommand{\mathi}{\mathrm{i}}

\newcommand{\Lagr}{\mathcal{L}}

\allowdisplaybreaks[1]
\newcounter{bla}

\journal{Computer Physics Communications}

\begin{document}

\begin{frontmatter}



\title{FeynArts model file for MSSM transition counterterms from DREG to
DRED}
\author{Dominik St\"ockinger\corref{cor1} and Philipp Var\v{s}o}
\address{Institut f\"ur Kern- und Teilchenphysik, TU Dresden, Dresden, Germany}
\cortext[cor1]{email: {\tt Dominik.Stoeckinger@tu-dresden.de}}

\begin{abstract}
The FeynArts model file \model\ implements MSSM  transition
counterterms which can convert one-loop Green functions from
dimensional regularization to dimensional 
reduction. They correspond to a slight extension of the well-known
Martin/Vaughn counterterms, specialized to the MSSM, and can serve
also as supersymmetry-restoring counterterms. The paper provides full
analytic results for  
the counterterms and gives one- and two-loop usage examples. The model
file can simplify combining \MSbar-parton distribution functions with 
supersymmetric renormalization or  avoiding the renormalization of
$\epsilon$-scalars in dimensional reduction. 
\end{abstract}

\end{frontmatter}

{\bf PROGRAM SUMMARY}

\begin{small}
\noindent
{\em Manuscript Title:} FeynArts model file for MSSM transition
counterterms from DREG to 
DRED                                       \\
{\em Authors:} Dominik St\"ockinger, Philipp Var\v{s}o         \\
{\em Program Title:}   \model                                       \\
{\em Licensing provisions:}  LGPL-License [1] \\
{\em Programming language:}  {\tt Mathematica, FeynArts}      \\
{\em Operating system:}     Any, with running  {\tt Mathematica, FeynArts} installation                                    
{\em Classification:}  4.4 	Feynman Diagrams,  5 Computer Algebra,
  11.1  General, High Energy Physics and Computing                                    \\
{\em Nature of problem:}
The computation of one-loop Feynman diagrams in the minimal
supersymmetric standard model (MSSM) requires regularization. Two schemes,
dimensional regularization and dimensional reduction are both common
but have different pros and cons. In order to combine the advantages
of both schemes one would like to easily convert existing results from
one scheme into the other.
   \\
{\em Solution method:}
Finite counterterms are constructed which correspond precisely to the
one-loop scheme differences for the MSSM. They are provided as  a
FeynArts [2] model file. Using 
this model file together with FeynArts, the (ultra-violet)
regularization of any MSSM one-loop Green function is switched
automatically from dimensional regularization to dimensional
reduction. In particular the counterterms serve as
supersymmetry-restoring counterterms for dimensional regularization. 
   \\
{\em Restrictions:}
The counterterms are restricted to the one-loop level and the MSSM.
   \\

\end{small}

\section{Introduction}

The minimal supersymmetric standard model (MSSM) is a
promising extension of the standard model. Many phenomenological
investigations require the computation of quantum corrections and
regularization of ultraviolet (UV) and infrared (IR) divergences. The
two most common regularization schemes for the 
MSSM are dimensional regularization (DREG) \cite{HV} and dimensional
reduction (DRED) \cite{Siegel79,CJN80}. In both schemes all momentum
integrals are formally continued to $D$ dimensions, allowing for very 
efficient integration techniques. In DREG also gauge fields are
treated as $D$-component quantities. This leads to a mismatch between
the number of degrees of freedom of gauge fields and gauginos ($D$
vs.\ 4) and the breaking of supersymmetry on the regularized level.
In DRED gauge fields remain formally  4-component quantities, and
therefore DRED is better suited for supersymmetric theories.

Nevertheless, as discussed e.g.\ in the report of the  ``Supersymmetry
Analysis Project''  \cite{SPA}, DREG should not be discarded as a
regularization of the MSSM. In particular, if hadronic processes
should be interfaced with the common \MSbar-parton distribution
functions, or if existing building blocks or algorithms in DREG should
be used, DREG can be
advantageous. Refs.\ \cite{BeenakkerHopker,QCDEWcorrections} provide
examples 
of MSSM computations where DREG has been used, and where the breaking
of supersymmetry has been compensated by adding appropriate
supersymmetry-restoring counterterms. Given that both DREG and DRED
have specific advantages it would be optimal to be able to combine
these advantages in practical computations.

In the present paper we introduce the model
file \model\ for the {\tt FeynArts} \cite{fa} package for generating Feynman
diagrams and amplitudes. The model file contains one-loop transition 
counterterms for the MSSM corresponding to switching the
UV-regularization from DREG to DRED. In particular it thus
automatically includes all necessary supersymmetry-restoring
counterterms for DREG. It is fully compatible to the original {\tt
  FeynArts} MSSM model file  and to further processing of the generated
amplitudes with {\tt FormCalc} \cite{fc} or similar programs.

The basic equation satisfied by the transition counterterm action
$\Gamma_{\rm ct, trans}^{(1)}$ implemented in our model file is 
\begin{equation}
\Gamma^{(1),\rm DRED}=\Gamma^{(1),\rm DREG}+\Gamma_{\rm
  ct,trans}^{(1)}+{\cal O}(D-4),
\label{transition}
\end{equation}
where $\Gamma^{(1),\rm RS}$ is the generating functional for one-loop
one-particle irreducible off-shell Green functions
regularized in the scheme RS. In  words, the transition 
counterterms translate off-shell (and IR-finite on-shell) Green
functions from DREG to DRED. The terms of ${\cal O}(D-4)$  are meant
to include Green functions with 
external so-called $\epsilon$-scalars, which exist only in DRED but
not in DREG, and which are discussed further below.
Analytical results for such transition
counterterms have already been published at the 
one-loop level for physical parameters in general supersymmetric
models in Ref.\ \cite{MartinVaughn} and at the two-loop level for
supersymmetric QCD in Ref.\ \cite{Mihaila}.

Since the IR-regularization is unaffected by the transition
(\ref{transition}), on-shell Green functions with IR  
divergences do not become equal by adding the transition
counterterms. This is desired since it makes possible to achieve UV
regularization by DRED and IR regularization by DREG or vice versa.
If e.g.\ hadronic MSSM processes are computed in this way, manifestly
supersymmetric UV renormalization can be easily combined with using
the customary \MSbar-parton distribution functions. 
The transition counterterms in our model file are thus complementary
to the transition rules presented in Ref.\ \cite{SignerDSnew}, which
correspond to switching the IR-regularization from DREG to DRED.

The outline of the present paper is as follows.  In the remainder of
this Introduction we briefly review the relevant status of DREG and
DRED. Section \ref{sec:usage} describes the installation and usage
of the model file. In section \ref{sec:computation} we explain the
theory behind the transition counterterms and collect analytical
results for generic supersymmetric theories. Section
\ref{sec:construction} is devoted to the specialization to the MSSM
and the implementation. In section \ref{sec:validation} we show which 
one- and two-loop tests we have carried out to validate the model
file. In section \ref{sec:conclusions} we conclude with further
remarks on possible applications. The Appendix contains the full
result of the MSSM transition counterterm Lagrangian.

In order to study DRED and its relation to DREG it is useful to
decompose the metric tensor appearing e.g.\ in the propagator
numerator of a regularized vector field as 
\begin{align}
\label{gghatgtilde}
g^{\mu\nu}&=\ghat^{\mu\nu}+\gtilde^{\mu\nu},
\end{align}
where $\ghat$ and $\gtilde$ are the metric tensors of the
$D$-dimensional and $(4-D)=2\epsilon$-dimensional subspaces. Accordingly, a
vector field $V^\mu$ can be decomposed into its $D$-dimensional and
$2\epsilon$-dimensional parts $\hat{V}^\mu=\ghat^{\mu\nu}V_\nu$,
$\tilde{V}^\mu=\gtilde^{\mu\nu}V_\nu$. $\hat{V}^\mu$ behaves as the
$D$-dimensional gauge field. $\tilde{V}^\mu$, on the other hand, has
the interactions and gauge transformations of scalar
fields in the adjoint representation with multiplicity $2\epsilon$,
hence the name $\epsilon$-scalars \cite{CJN80}. 
Several subtle problems of DRED have been reported in the literature,
most notably Siegel's inconsistency \cite{Siegel80}, the violation of
unitarity \cite{HooftvanDamme}, and the violation of infrared
factorization \cite{BeenakkerSmith}. These problems are
reviewed and stressed e.g.\ in
Refs.\ \cite{JJ,SPA}.  In recent 
years, significant 
  progress in the understanding of DRED has been achieved in all
  desired directions. The current status can be summarized as
  follows:
\begin{itemize}
\item DREG and DRED can both be formulated in a mathematically
  consistent way, such that any calculation leads to an unambiguous
  answer \cite{BM,DS05}. The consistent formulations justify the
  required formally $D$- and 4-dimensional algebraic operations such
  as $g^{\mu}{}_\mu=4$. But in DRED they forbid to use certain
  strictly 4-dimensional identities 
  related to assuming that the l.h.s.\ of Eq.\ (\ref{gghatgtilde}) has
  the explicit form diag($1,-1,-1,-1$), thus avoiding Siegel's
  inconsistency \cite{Siegel80}. 
\item DREG and DRED are equivalent, i.e.\ for any theory regularized
  in DREG with certain bare parameters there is a corresponding theory
  regularized in DRED with suitably chosen bare parameters and fields
  for which the S-matrix and Green functions (ignoring IR divergences)
  are equal \cite{JJR}. This theorem proves the existence of
  transition counterterms like in Eq.\ (\ref{transition}) at all
  orders for any theory. It is only valid if the masses and couplings
  of $\epsilon$-scalars are renormalized independently, but
  Ref.\ \cite{JJR} also shows that the renormalized $\epsilon$-scalar
  masses and couplings can be chosen at will. It implies that
  DRED applied in this way preserves unitarity.\footnote{%
The version of DRED used in Ref.\ \cite{JJR} corresponds to the
consistent version of Ref.\ \cite{DS05} since the multiplicity of the
$\epsilon$-scalars is kept arbitrary throughout the computation,
which corresponds to the ``quasi-4-dimensional'' treatment of
Ref.\ \cite{DS05}.}
\footnote{For recent explicit examples of the need
    for an independent renormalization of $\epsilon$-scalars see
  Refs.\ \cite{SteinhauserKantPaper1,Kant:2010tf,Kilgore}. For an
  attempt to rescue a renormalization scheme with equal
  treatment of gauge fields and $\epsilon$-scalars see
  Ref.\ \cite{Boughezal:2011br}.} 
\item At the one-loop level DRED preserves supersymmetry. For a list
  of references and methods and recent results for higher orders, see
  e.g.\ \cite{JJ,DS05,DREDProc,HollikDS05,Hermann:2011ha}. 
\item At the one-loop level DREG and DRED are both compatible with QCD
  factorization of IR singularities. In DRED, the $\epsilon$-scalar
  gluons have to be treated as independent QCD partons, which
  contribute to splitting functions of other partons and have their
  own splitting functions. The resulting difference between
  (UV-renormalized but IR divergent) DREG and DRED virtual and real
  next-to-leading 
  order corrections can be cast into simple transition rules
  \cite{Signer:2005iu,SignerDSnew}.\footnote{%
When comparing Refs.\ \cite{SignerDSnew} and \cite{Kilgore} one should
note that the term ``four-dimensional helicity scheme'' (FDH) is used
differently in both references. In the former reference
it denotes a regularization scheme which differs slightly from DRED,
while in the latter reference it implies a certain renormalization
prescription which leads to incorrect results. The modifications to FDH
proposed in the Conclusions of Ref.\ \cite{Kilgore} are in fact in
agreement with the way renormalization has been done in
Ref.\ \cite{SignerDSnew}.}
\end{itemize}

\section{Installation and usage of the model file}
\label{sec:usage}

In order to use our model file, two files need to be copied into
the {\tt Models/} directory of a complete {\tt FeynArts} installation:
\begin{itemize}
\item {\tt LorentzTadpole.gen}: a replacement generic model file which
  differs from the original {\tt Lorentz.gen} only by the possibility
  of tadpole (one-point) counterterm Feynman rules. 
\item \model: the model file containing all the transition
  counterterms. It is realized as an Add-on model file, building on the
  original {\tt MSSMQCD.mod} model file.
\end{itemize}
The two files can be obtained from the web page {\tt http://iktp.tu-dresden.de/?theory-software}
either separately or in a {\tt .tar} archive, together with
documentation. 

The model file is used just like any other, by the rule {\tt
  Model->\modelname}, which must be accompanied by {\tt
  GenericModel->LorentzTadpole} when using the {\tt FeynArts
  InsertFields} function. 
Since it uses the same naming conventions as the original {\tt
  MSSMQCD} model file it is fully compatible not only with {\tt
  FeynArts} but also with further programs such as {\tt FormCalc} or
{\tt TwoCalc}.

We now give two code examples for a typical use of the model file,
corresponding to the two validation tests described below in section
\ref{sec:validation}. 
First, the contribution of the electron self energy to
Eq.\ (\ref{transition}) can
be computed as follows in a {\tt 
  Mathematica} session where {\tt FeynArts} and {\tt FormCalc} are
loaded. The one-loop self energy and counterterm diagrams are generated by 
\begin{verbatim}
top1L = CreateTopologies[1, 1->1,
             ExcludeTopologies->{Tadpoles, Internal}];
topCT = CreateCTTopologies[1, 1 -> 1,
             ExcludeTopologies -> {Tadpoles, Internal}];
amp1L = CreateFeynAmp[InsertFields[top1L, F[2,{1}]->F[2,{1}],
             Model -> MSSMdreg2dred, GenericModel -> LorentzTadpole]];
ampCT = CreateFeynAmp[InsertFields[topCT, F[2,{1}]->F[2,{1}],
             Model -> MSSMdreg2dred, GenericModel -> LorentzTadpole]];
\end{verbatim}
Note that the same model file can be used for the one-loop and
counterterm diagrams. For {\tt amp1L} the choice {\tt Model->MSSM}
would lead to the same result. With these definitions we can compute
\begin{verbatim}
GammaDRED = Plus @@ CalcFeynAmp[amp1L, OnShell -> False, 
             Dimension -> 4, FermionChains -> Dirac] //. Abbr[]; 
GammaDREG = Plus @@ CalcFeynAmp[amp1L, OnShell -> False,
             Dimension -> D, FermionChains -> Dirac] //. Abbr[]; 
GammaCTtrans = Plus @@ CalcFeynAmp[ampCT, OnShell -> False,
             Dimension -> D, FermionChains -> Dirac] //. Abbr[]; 
\end{verbatim}
After identifying {\tt ME} with {\tt Mf[2, 1]} we obtain that {\tt
  GammaDRED=GammaDREG+GammaCTtrans}, as prescribed by
Eq.\ (\ref{transition}). 

As a second example, the one-loop counterterm diagrams for the
two-loop selectron self energy such as the one in
Fig.\ \ref{fig:twoloopdiagrams} are generated by  
\begin{verbatim}
InsertFields[
   CreateCTTopologies[2, 1 -> 1, ExcludeTopologies ->
                     {WFCorrections, Tadpoles, TadpoleCTs, Internal}],
   S[12,{1,1}] -> S[12,{1,1}],
   InsertionLevel -> Particles,
   Model -> MSSMdreg2dred, 
   GenericModel -> LorentzTadpole
  ];
\end{verbatim}
in a {\tt Mathematica} session where {\tt FeynArts} is loaded.
Section \ref{sec:validation} below discusses the result for the
two-loop selectron self energy.

\section{Computation and results for a generic theory}
\label{sec:computation}

We begin with the transition counterterms in a generic softly broken
supersymmetric gauge theory with simple gauge group. Using the same
conventions as 
Ref.\ \cite{MartinVaughn}, the gauge fields, gauginos and chiral
supermultiplets are denoted as $V_a^\mu$, $\lambda_a$, $\Phi_i$. The
chiral supermultiplets consist of scalar fields $\phi_i$ and
two-component Weyl fermions $\psi_i$ and transform under 
representations $r_i$ under the gauge group. The generators
$T^a_{(i)}$ for the representation $r_i$ satisfy $\sum_a
(T^a_{(i)}T^a_{(i)})_{k}{}^l=C(r_i)\delta_{k}{}^l$, and the structure
constants of the gauge group satisfy
$f_{abc}f_{dbc}=C(G)\delta_{ad}$. The superpotential is given
by $W=\frac16 Y^{ijk}\Phi_i\Phi_j\Phi_k$, and the soft breaking
gaugino mass term by ${\cal L}_{\rm soft}=-\frac12
m_\lambda\lambda_a\lambda_a+$h.c.. Further soft breaking terms turn
out to be irrelevant for the analysis. The theory is quantized in the
usual $R_\xi$-gauges, where $\epsilon$-scalars do not appear in the
gauge fixing  and ghost terms.

\begin{figure}
\centerline{
\null
\hfill
\begin{picture}(130,90)(0,-20)
\SetWidth{.8}
\Text(15,25)[t]{$\psi$}
\Text(105,25)[t]{$\psi$}
\ArrowLine(0,30)(30,30)
\ArrowLine(90,30)(120,30)
\ArrowArc(60,30)(30,180,360)
\PhotonArc(60,30)(30,0,180){2}{10.5}
\Vertex(30,30){2}
\Vertex(90,30){2}
\Text(60,-10)[t]{(a)}
\end{picture}
\hfill
\begin{picture}(130,90)(0,-20)
\SetWidth{.8}
\Text(15,25)[t]{$V$}
\Text(105,25)[t]{$V$}
\Photon(0,30)(30,30){2}{4}
\Photon(90,30)(120,30){2}{4}
\PhotonArc(60,30)(30,180,360){2}{10.5}
\PhotonArc(60,30)(30,0,180){2}{10.5}
\Vertex(30,30){2}
\Vertex(90,30){2}
\Text(60,-10)[t]{(b)}
\end{picture}
\hfill
\begin{picture}(130,90)(0,-20)
\SetWidth{.8}
\Text(15,25)[t]{$\phi$}
\Text(105,25)[t]{$\phi$}
\DashArrowLine(0,30)(30,30){4}
\DashArrowLine(90,30)(120,30){4}
\DashArrowArc(60,30)(30,180,360){4}
\PhotonArc(60,30)(30,0,180){2}{10.5}
\Vertex(30,30){2}
\Vertex(90,30){2}
\Text(60,-10)[t]{(c)}
\end{picture}
\hfill
}
\caption{
\label{fig:samplediff}
Three diagrams illustrating the difference between DREG and
DRED. Wavy lines denote vector fields (i.e.\ gauge fields and
$\epsilon$-scalars in DRED); solid and dashed lines denote
the fermionic and scalar components of a 
chiral supermultiplet. 
Diagrams (a) and (b) are different in DREG and DRED, diagram (c) is
equal in both regularizations.}
\end{figure}
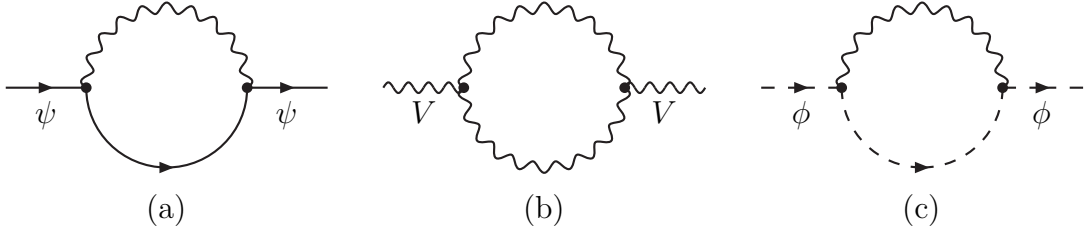

The computation of the transition counterterms differs from the ones in
Refs.\ \cite{MartinVaughn,Mihaila} in that we require equality of
all off-shell one-loop Green functions, not only of physical
quantities. Fig.\ \ref{fig:samplediff} shows sample one-loop diagrams
which illustrate the structure of the regularization dependence. The
only difference between DREG and DRED originates from the additional
degrees of freedom of the vector fields in DRED, the
$\epsilon$-scalars. In diagrams (a) and (b) the $\epsilon$-scalars
give contributions to the numerator algebra of the order
$\epsilon=(4-D)/2$, which combine with the $\frac{1}{\epsilon}$ poles
of the loop integral to a finite difference between DREG and DRED. In
diagram (c) the scalar--vector coupling is proportional to momenta,
which are always regularized in $D$ dimensions, and hence the
$\epsilon$-scalars cannot contribute and there is no difference.

We have computed the difference between DREG and DRED for all one-loop
one-particle irreducible Green functions and thus determined the
transition counterterms, defined by Eq.\ (\ref{transition}) for the
generic theory. They can be written as 
\begin{align}
\Gamma_{\rm ct, trans}^{(1)} &= \int d^4x {\cal L}_{\rm ct,trans}
\end{align}
where the counterterm Lagrangian ${\cal L}_{\rm ct,trans}$ can be
obtained from the original Lagrangian by suitable field and parameter
renormalizations. We find full agreement with
Ref.\ \cite{MartinVaughn} for the parameter renormalization, while the
field renormalization transition counterterms have not been published
before. For completeness and convenience we list all results in the
following. 
\begin{itemize}
\item The parameter counterterms respect gauge
  invariance but not supersymmetry and cannot be directly obtained by
  multiplicative renormalization of the original
  Lagrangian. The only parameters of the original Lagrangian are $g$,
  $Y^{ijk}$, $m_\lambda$ and further soft breaking 
  and gauge fixing parameters. As shown
  in \cite{MartinVaughn} the transition counterterms require us to
  distinguish the actual gauge coupling $g$, which appears in all
  couplings to gauge fields, from the couplings $\ghat_i$, which appear
  in the gaugino interactions with $\phi_i$ and $\psi_i$, and by treating
  the Yukawa couplings $Y^{ijk}\phi_i\psi_j\psi_k$ as non-symmetric
  in $(ijk)$. The transition counterterms can then be applied by
  renormalizing these parameters multiplicatively, in the
  form $p\to (1+\delta Z_p)p$, with
\begin{subequations}
\label{transpar1}
\begin{align}
\delta Z_g&= -\frac{g^2}{96\pi^2}C(G)\\
\delta Z_{\ghat_i}&=\frac{g^2}{32\pi^2}\left(C(G)-C(r_i)\right)\\
\delta Z_{Y^{ijk}}&=\frac{g^2}{32\pi^2}\left(C(r_j)+C(r_k)-2C(r_i)\right)\\
\delta Z_{m_\lambda}&=\frac{g^2}{16\pi^2}C(G)
\end{align}
\end{subequations}
The quartic scalar interactions cannot be treated in a multiplicative
way. Instead, we need to add the counterterm Lagrangian
$-\frac14\delta\lambda_{ij}{}^{kl}\phi^*_i\phi^*_j\phi_k\phi_l$ with
\begin{align}
\label{translambda}
\delta\lambda_{ij}{}^{kl} &=-\frac{g^4}{16\pi^2}
\{T^a,T^b\}_i{}^k  \{T^a,T^b\}_j{}^l + (i\leftrightarrow j).
\end{align}
Here $T^a$ denotes the block matrices for the generators of the full,
reducible representation consisting of all irreducible representations
$r_i$. No counterterms corresponding to gauge fixing or soft breaking
scalar mass parameters are required.
\item The field renormalization counterterms arise from applying the
  renormalization transformation $\phi\to(1+\frac12\delta Z_\phi)\phi$
  to all fields with
\begin{subequations}
\label{transfield}
\begin{align}
\delta Z_V&=\frac{g^2}{48\pi^2}C(G)\\
\delta Z_\lambda&=\frac{g^2}{16\pi^2}C(G)\\
\delta Z_{\psi_i}&=\frac{g^2}{16\pi^2}C(r_i)
\end{align}
\end{subequations}
on the original gauge invariant Lagrangian of the generic theory. 
No such
renormalizations are required for scalar or ghost fields. The
field renormalization counterterms do not modify physical quantities
but are required to obtain equality between Green functions. 
\end{itemize}
Note
that the simple abelian-like relation $\delta Z_g+\frac12\delta Z_V
=0$ holds since $\epsilon$-scalars do not interact with gauge fixing
and ghost terms. For the same reason, gauge fixing
  and ghost terms do not require transition
  counterterms, and the
  renormalization transformations must not be applied to gauge fixing
  and ghost terms.

\begin{figure}
\centerline{
\null
\hfill
\begin{picture}(130,90)(0,0)
\SetWidth{.8}
\Text(15,25)[t]{$\epsilon$}
\Text(105,25)[t]{$\epsilon$}
\Text(30,75)[b]{$\phi$}
\DashLine(0,30)(60,30){1}
\DashLine(60,30)(120,30){1}
\DashArrowArc(60,60)(30,0,360){4}
\Vertex(60,30){2}
\Text(60,15)[t]{(a)}
\end{picture}
\hfill
\begin{picture}(130,90)(0,0)
\SetWidth{.8}
\Text(15,25)[t]{$\phi$}
\Text(105,25)[t]{$\phi$}
\Text(30,75)[b]{$\epsilon$}
\DashArrowLine(0,30)(60,30){4}
\DashArrowLine(60,30)(120,30){4}
\DashCArc(60,60)(30,0,360){1}
\Vertex(60,30){2}
\Text(60,15)[t]{(b)}
\end{picture}
\hfill
}
\caption{
\label{fig:subtle}
Two diagrams illustrating the independent role of
$\epsilon$-scalars. Dashed lines denote the scalar component $\phi$ of a
chiral supermultiplet; dotted lines denote $\epsilon$-scalars. 
(a) Scalar contribution to the $\epsilon$-scalar self energy, which
requires renormalization of the $\epsilon$-scalar mass;
(b) $\epsilon$-scalar contribution to a scalar self energy, which
depends on the tree-level $\epsilon$-scalar mass.}
\end{figure}
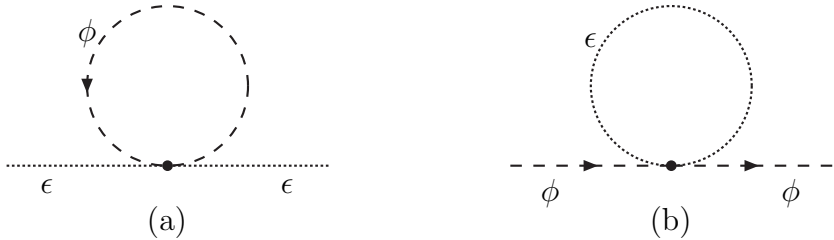

We close the section with a reminder of several subtleties related
to the $\epsilon$-scalars in the equivalence theorem between DREG and
DRED \cite{JJR}. First, it is clearly impossible to define transition
counterterms for Green functions with external $\epsilon$-scalars. In
DREG there are no $\epsilon$-scalars, and in DRED the best we can do
is to renormalize Green functions with external $\epsilon$-scalars
such that they become finite. At the one-loop level Green functions
with  external $\epsilon$-scalars can be ignored if we do not desire
to regularize IR singularities in DRED \cite{SignerDSnew}. However, if
Green functions with external $\epsilon$-scalars appear 
as subgraphs in higher-order graphs their renormalization is vital in
order to obtain results consistent with renormalizability, unitarity,
and equivalence to DREG 
\cite{JJR,SteinhauserKantPaper1,Kilgore}.  In theories with softly
broken supersymmetry, the diagram in Fig.\ \ref{fig:subtle}(a)
produces a divergent contribution which can only be cancelled by
renormalization of an $\epsilon$-scalar mass. It 
thus exemplifies that in DRED, $\epsilon$-scalar masses (and in
general also couplings) need to be renormalized independently of the
corresponding gauge field parameters.  Diagram (b) is an example
of a one-loop diagram whose finite part depends on the tree-level
$\epsilon$-scalar mass in DRED. As shown in \cite{JJR}, all choices of
the $\epsilon$-scalar mass lead to equivalent results, and it is no
restriction to set it to zero at the renormalized level, as done
e.g.\ in the \DRbar' scheme \cite{Jack:1994rk}. 

The transition counterterms listed in the present section have been
evaluated for zero tree-level $\epsilon$-scalar mass. They are
sufficient to satisfy Eq.\ (\ref{transition}), and they form the basis
of our {\tt FeynArts} model file. $\epsilon$-scalar mass counterterms
will be required for a consistent multi-loop renormalization of the
MSSM in DRED, but these are not the focus of the present paper.

\section{Specialization to the MSSM and the model file}
\label{sec:construction}

The MSSM is a softly broken supersymmetric
SU(3)$\times$SU(2)$\times$U(1) gauge theory with chiral superfields
for quarks, leptons and two Higgs doublets. The generic result for the
transition counterterms can be specialized to the MSSM after applying
two generalizations \cite{MartinVaughn}. First, the three gauge group
factors $G_{1,2,3}$ with gauge couplings $g_{1,2,3}$ can be taken into
account by applying the rules of Ref.\ \cite{Machacek:1983tz},
i.e.\ by replacing $g^2 C(G)\to g_a^2 C(G_a)$, $g^2 C(r_i)\to \sum_b
g_b^2 C_b(r_i)$ in Eqs.\ (\ref{transfield},\ref{transpar1}), and
replacing $g T^a$ by $g_r T^a_{(r)}$ in Eq.\ (\ref{translambda}),
where $T^a_{(r)}$ is the $G_r$ generator, and carrying out a double
sum over the gauge groups. Second, the dimensionful $\mu$-parameter in
the MSSM superpotential can be treated like a Yukawa coupling to a
spurion field for which $C(r)=0$. 

We have written a {\tt Mathematica} program which implements all these
rules and generates the full transition counterterm Lagrangian for a
generic model, and we have specialized this program to the MSSM. Using
this program we have obtained the MSSM transition counterterm
Lagrangian in interaction eigenstates, in a form appropriate for input
to the package {\tt FeynRules} \cite{fr}. This Lagrangian is explicitly
reprinted in  \ref{app:transition}. 

As a second step we have implemented the spontaneous symmetry breaking
and the mixing of interaction to mass eigenstate fields in the MSSM as
an input file for {\tt FeynRules}. Here we have used the same
conventions for mixing matrices as the original {\tt FeynArts} MSSM
model file \cite{fa}, and we have followed the restriction to neglect
family mixing in the sfermion sector. In contrast to the original
model file, we also neglect family mixing in the fermion sector,
i.e.\ we set the CKM matrix to the unit matrix. Then we have used {\tt
  FeynRules} to generate the transition counterterm Lagrangian
expressed in terms of mass eigenstates.

After spontaneous symmetry breaking and inserting Higgs vacuum
expectation values, 3- and 4-point interactions involving Higgs fields
generate also 2- and 1-point interactions. Because of this, the MSSM
transition counterterm Lagrangian contains not only gaugino but also
fermion, vector and scalar self energy transition counterterms, as
well as scalar tadpole transition counterterms. A complete transition
from DREG to DRED requires all these counterterms. Even in a
renormalization scheme where tadpoles are renormalized to zero, 
tadpole graphs and hence tadpole transition counterterms are needed as
they contribute to self energy or other counterterm insertions.

In order to take these necessary counterterms into account, several
modifications to {\tt FeynRules} and {\tt FeynArts} had to be made.
\begin{itemize}
\item The possibility to generate Feynman rules for self energy and
  tadpole counterterms has been implemented in {\tt FeynRules}. 
\item The possibility to allow tadpole counterterms and generate
  tadpole counterterm diagrams has been implemented in {\tt FeynArts}
  in terms of a replacement for the generic {\tt Lorentz.gen} model
  file. We call our generic model file {\tt LorentzTadpole.gen}.
\end{itemize}
Finally, with this procedure we have generated the {\tt FeynArts} 
model file \model\ for all the transition counterterms of the MSSM. It
has the form of an Add-on model file, and it has to be used together
with our {\tt  LorentzTadpole.gen} generic model file. The usage has been
described in section \ref{sec:usage}.
A lengthy collection of
transition Feynman rules contained in this model 
file can obtained from the web page {\tt
  http://iktp.tu-dresden.de/?theory-software}.

\section{Validation}
\label{sec:validation}

We have validated our model file in two ways. As a first and direct
test we have evaluated the contribution to Eq.\ (\ref{transition}) of {\em
  every} MSSM Green function for which transition counterterms
exist. The computation is analogous to the one described in section
\ref{sec:usage} for the electron self energy, suitably generalized and
automatized. On the one hand, {\tt FeynArts/FormCalc} was used to
compute the one-loop contributions both in 
DREG and DRED, and the
difference between DREG and DRED in the limit $D\to4$ was
evaluated. On the other hand, the corresponding contributions of the
transition counterterms were computed using {\tt FeynArts/FormCalc}
with our transition counterterm model file. We found full agreement,
thus explicitly verifying Eq.\ (\ref{transition}). The source code of
this program, {\tt validation.m}, can be obtained from the web
page {\tt http://iktp.tu-dresden.de/?theory-software}. It is easily
adaptable, provides a 
comprehensive check and illustrates the usage further. It requires
only a {\tt FeynArts} and {\tt FormCalc} installation.

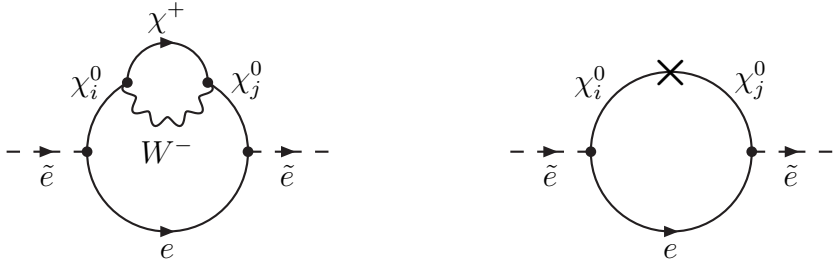
\begin{figure}
\centerline{
\null
\hfill
\begin{picture}(130,90)(0,0)
\SetWidth{.8}
\Text(15,25)[t]{$\tilde{e}$}
\Text(105,25)[t]{$\tilde{e}$}
\Text(60,75)[b]{$\chi^+$}
\Text(60,35)[t]{$W^-$}
\Text(60,-5)[t]{$e$}
\Text(30,50)[b]{$\chi^0_i$}
\Text(90,50)[b]{$\chi^0_j$}
\DashArrowLine(0,30)(30,30){4}
\DashArrowLine(90,30)(120,30){4}
\CArc(60,30)(30,0,60)
\CArc(60,30)(30,120,180)
\ArrowArc(60,30)(30,180,360)
\PhotonArc(60,56)(15,180,360){2}{5.5}
\ArrowArcn(60,56)(15,180,360)
\Vertex(30,30){2}
\Vertex(90,30){2}
\Vertex(45,56){2}
\Vertex(75,56){2}
\end{picture}
\hfill
\begin{picture}(130,90)(0,0)
\SetWidth{.8}
\Text(15,25)[t]{$\tilde{e}$}
\Text(105,25)[t]{$\tilde{e}$}
\Text(60,-5)[t]{$e$}
\Text(30,50)[b]{$\chi^0_i$}
\Text(90,50)[b]{$\chi^0_j$}
\DashArrowLine(0,30)(30,30){4}
\DashArrowLine(90,30)(120,30){4}
\CArc(60,30)(30,0,90)
\CArc(60,30)(30,90,180)
\ArrowArc(60,30)(30,180,360)
\Vertex(30,30){2}
\Vertex(90,30){2}
\Text(60,60)[c]{{\bf \Large\boldmath{$\times$}}}
\end{picture}
\hfill
}
\caption{
\label{fig:twoloopdiagrams}
Sample two-loop diagram for the selectron self energy for which the
DREG and DRED results differ by more than a polynomial of the form
$(a+ b p^2)$, and a corresponding one-loop counterterm diagram with
insertion of a transition counterterm.}
\end{figure}

As a second, less comprehensive but more intricate test we have
considered an MSSM two-loop computation with renormalization of the
one-loop subdiagrams. We have used {\tt FeynArts/TwoCalc} \cite{2lred}
to compute the selectron self energy
$\Gamma_{\tilde{e}\tilde{e}^\dagger}^{(2)}$ at the two-loop level in DREG and
DRED. For simplicity we have carried out the calculation numerically,
for an MSSM parameter point where all mixing matrices are non-trivial,
and we have expanded in the external momentum being small. It turned
out that the dependence of the difference on 
the external momentum was {\em not} a polynomial of second
degree. Fig.\ \ref{fig:twoloopdiagrams}(a)  shows a sample diagram
which contributes in this way. However, after adding one-loop diagrams
$\Gamma_{\tilde{e}\tilde{e}^\dagger}^{(1+{\rm ct,DREG})}$ with insertions
of the one-loop transition counterterms, see
Fig.\ \ref{fig:twoloopdiagrams}(b), we obtain that 
\begin{equation}
\Gamma_{\tilde{e}\tilde{e}^\dagger}^{(2,\rm DRED)}(p^2)  =
\Gamma_{\tilde{e}\tilde{e}^\dagger}^{(2,\rm DREG)}(p^2)  +
\Gamma_{\tilde{e}\tilde{e}^\dagger}^{(1+{\rm ct},\rm DREG)}(p^2)
+a+b p^2+{\cal O}(D-4).
\end{equation}
Here, the  polynomial $(a+b p^2)$ is of a form that could be absorbed
by a local mass and field renormalization counterterm, corresponding
to a two-loop transition counterterm $\Gamma_{\rm ct,trans}^{(2)}$. This is exactly
what is expected from the general statement that DREG and DRED are
equivalent and one can find transition counterterms at all
orders. Since many transition counterterms contribute to
$\Gamma_{\tilde{e}\tilde{e}^\dagger}^{(1+{\rm ct},\rm DREG)}(p^2)$, and
since the two-loop diagrams involve up to $1/(D-4)^2$ divergences, this
constitutes a non-trivial test of our counterterm model
file.\footnote{%
In principle, such a test  requires a subloop renormalization
in DRED, rendering subdiagrams involving external $\epsilon$-scalars
finite. In the MSSM this amounts to  adding one-loop counterterm
diagrams with appropriate $\epsilon$-scalar mass counterterm
insertions to the DRED result 
$\Gamma_{\tilde{e}\tilde{e}^\dagger}^{(2,\rm DRED)}(p^2)$.  However, in
the case of the selectron self energy these extra counterterm diagrams
do not contribute since there is no
selectron--selectron--$\epsilon$-scalar coupling and since we set the
tree-level $\epsilon$-scalar mass to zero.}

\section{Conclusions}
\label{sec:conclusions}

In the present paper we have introduced a {\tt FeynArts} model file 
\model, which implements MSSM one-loop transition counterterms from
DREG to DRED. This model file in particular contains all required
supersymmetry-restoring counterterms for DREG, and it thus allows to
use DREG for MSSM loop calculations without violating supersymmetry. 
It allows to easily compare calculations, or to combine building
blocks computed in different regularization schemes.

The present paper also contains complete results for the
transition counterterms both in a generic supersymmetric model and in
the MSSM, including field renormalization counterterms. We have
shown that the gauge fixing and ghost sectors do not require such
counterterms. These results and the model file should be valuable for
a detailed theoretical understanding of DREG and DRED and as tools for
practical calculations. 

There are two main areas where the use of the transition counterterm
model file can simplify calculations.
One is the computation of
MSSM processes in the presence of IR divergent QCD corrections. 
Here it is advantageous to regularize the IR divergences in DREG, both
because of the existing \MSbar-parton distribution functions (PDFs) and the
complicated structure of factorization in DRED
\cite{BeenakkerSmith,SignerDSnew,Signer:2005iu}.   
In the
past, e.g.\ in Refs.\ \cite{BeenakkerHopker,QCDEWcorrections}, DREG
has been used for the supersymmetric QCD-part of the calculations, so
that \MSbar-PDFs could be used, and the required
supersymmetry-restoring counterterm \cite{MartinVaughn} has been added
by hand. This procedure required a mixed DRED/DREG
regularization in Ref.\ \cite{QCDEWcorrections}, where also
electroweak corrections were taken into account.

With our model file at hand, one can now regularize the full MSSM
process in DREG and then automatically add the transition
counterterms. In this way UV-regularization by DRED and
IR-regularization by DREG are combined in a straightforward
way. Supersymmetry is manifest and e.g.\ the \DRbar-definition of MSSM 
parameters can be used, and simultaneously the IR divergences have the
usual, simple DREG-structure, \MSbar-PDFs can be employed and real
corrections can be simply computed in DREG.

A second area where the transition counterterm model file can be
helpful are multi-loop calculations which would require the renormalization of
$\epsilon$-scalars in DRED. As mentioned above, even though DRED
preserves supersymmetry, $\epsilon$-scalar masses have to be
renormalized independently in order to guarantee correct results
beyond the one-loop level. For a recent three-loop example and a
description of the subtleties involved see
Ref.\ \cite{Kant:2010tf}. If instead DREG is used together with the
transition counterterms, the correct DRED result corresponding to zero
renormalized $\epsilon$-scalar mass is directly obtained. To the
extent that no genuine two-loop transition counterterms are needed
this constitutes a potentially simpler alternative procedure.

\section*{Acknowledgements}

We thank Alexander Voigt for helpful discussions and further tests of
the model file.

\begin{appendix}

\section{MSSM transition counterterm Lagrangian}
\label{app:transition}

In the following we provide the explicit form of the MSSM transition
counterterm Lagrangian, expressed in terms of interaction eigenstate
fields.

The MSSM chiral superfields are denoted as $Q_{f,i,c}$,
$\bar{u}_{f,c}$, $\bar{d}_{f,c}$,  $l_{f,i}$, $\bar{e}_i$,
${H_{d}}_{i} $, ${H_{u}}_i$, where $f, i,c$ are family indices,
SU(2) dublet indices and colour indices, respectively. The scalar and
fermionic components are denoted as $\phi$ and $\Psi$, with
appropriate indices. The vector fields and gauginos are denoted as
$V_\mu$ and $\tilde\lambda$, with indices $G,W,B$ for SU(3), SU(2),
U(1), respectively, and an adjoint group index $a$, if
appropriate. $T^a$ and $\tau^a/2$ are the fundamental SU(3) and SU(2)
generators, and $f^{abc}$ and $\varepsilon^{abc}$ are the structure
constants. The gauge and family-diagonal Yukawa couplings are denoted
as $g_{3,2,1}$, $y^{e,d,u}_{f_1\, f_2}$, and the gaugino masses as
$m_{\lambda_{G,W,B}}$.  
\label{sec:full-MSSM-CT-Lag}
\begin{align}
  \Lagr_{\rm ct,trans}^{\rm MSSM} &= \restLL{\text{2-pt}} +   \restLL{\text{gaugino}}  + \restLL{\text{gauge}} + \restLL{\text{Yukawa}} + \restLL{\text{quartic}} 
\end{align}
\begin{align*}
  \restLL{\text{2-pt}}
  &= 
   \frac{g_2^2}{16 \pi ^2}
   \left(
     \mathi \barGaugino{W}^a \gamma^{\mu}
     \partial_{\mu} \Gaugino{W}^{a} -
     2 m_{\gaugino{W}} \barGaugino{W}^{a} \pleft \Gaugino{W}^{a}
     - 2m_{\gaugino{W}}^{\ast} \barGaugino{W}^{a} \pright \Gaugino{W}^{a}
   \right)
   \\
   &\hphantom{{}=} 
   + \frac{3 g_3^2}{32 \pi ^2}
   \left(
     \mathi \barGaugino{G}^a \gamma^{\mu}
     \partial_{\mu} \Gaugino{G}^{a} -
     2 m_{\gaugino{G}} \barGaugino{G}^{a} \pleft \Gaugino{G}^{a}
     - 2 m_{\gaugino{G}}^{\ast} \barGaugino{G}^{a} \pright \Gaugino{G}^{a}
   \right)
   \\
   &\hphantom{{}=} 
   + \frac{\mathi \left(g_1^2+27 g_2^2+48 g_3^2\right)}{576 \pi ^2}
   \left(
     \barFermion{Q}_{f_1,i_1,c_1} \gamma_{\mu}
     \partial^{\mu} \pleft \Fermion{Q}_{f_1,i_1,c_1}
   \right)
   \\
   &\hphantom{{}=} 
   + \frac{\mathi \left(g_1^2+3 g_2^2\right)}{64 \pi ^2}
   \left(
     \barFermion{l}_{f_1,i_1} \gamma_{\mu}
     \partial^{\mu} \pleft \Fermion{l}_{f_1,i_1}
   \right)
   \\
   &\hphantom{{}=} 
   + \frac{\mathi \left(g_1^2+3 g_2^2\right)}{64 \pi ^2}
   \left(
     \barFermion{H_u}_{i_1} \gamma_{\mu}
     \partial^{\mu} \pleft \Fermion{H_u}_{i_1}
   \right)
   \\
   &\hphantom{{}=}
   + \frac{\mathi \left(g_1^2+3 g_2^2\right)}{64 \pi ^2}
   \left(
     \barFermion{H_d}_{i_1} \gamma_{\mu}
     \partial^{\mu} \pleft \Fermion{H_d}_{i_1}
   \right)
   \\
   &\hphantom{{}=}
   + \frac{\mathi \left(g_1^2+3 g_3^2\right)}{36 \pi ^2}
   \left(
     \barCCFermion{u}_{f_1,c_1} \gamma_{\mu}
     \partial^{\mu} \pleft \CCFermion{u}_{f_1,c_1}
   \right)
   \\
   &\hphantom{{}=}
   + \frac{\mathi \left(g_1^2+12 g_3^2\right)}{144 \pi ^2}
   \left(
     \barCCFermion{d}_{f_1,c_1} \gamma_{\mu}
     \partial^{\mu} \pleft \CCFermion{d}_{f_1,c_1}
   \right)
   \\
   &\hphantom{{}=}
   + \frac{\mathi g_1^2}{16 \pi ^2}
   \left(
     \barCCFermion{e}_{f_1} \gamma_{\mu}
     \partial^{\mu} \pleft \CCFermion{e}_{f_1}
   \right)
   \\
   &\hphantom{{}=}
   + \frac{g_1^2+3 g_2^2}{32 \pi ^2}
   \left(
     \mu \, \epsilon_{i_1i_2} (\barFermion{{H_d}})^{\chargec}_{i_1}
     \pleft \Fermion{{H_u}}_{i_2} +
     \mu^{\ast} \epsilon^{\dagger}_{i_1i_2} \barFermion{{H_u}}_{i_2} 
     \pright (\Fermion{{H_d}})^{\chargec}_{i_1}
   \right)
   \\
   &\hphantom{{}=}
   -\frac{g_2^2}{48 \pi ^2}
   \left(
     (\partial_{\mu} {V_{W}^{a}}_{\nu})(\partial^{\mu} V_{W}^{\nu,a})
     - (\partial^{\mu} {V_{W}^{a}}_{\mu})(\partial^{\nu} {V_{W}^{a}}_{\nu})
   \right)
   \\
   &\hphantom{{}=}
   -\frac{g_3^2}{32 \pi ^2}
   \left(
     (\partial_{\mu} {V_G^a}_\nu)(\partial^{\mu} V_{G}^{\nu,a})
     - (\partial^{\mu} {V_{G}^{a}}_{\mu})(\partial^{\nu} {V_{G}^{a}}_{\nu})
   \right)
\end{align*}
\begin{align*}
  \restLL{\text{gaugino}} 
  &=
  -\frac{g_2^3}{4 \sqrt{2} \pi ^2}
  \left(
    \barFermion{l}_{f_1,i_1} \pright
    \Gaugino{W}^{a} \scalar{l}_{f_1,i_2} +
    \barGaugino{W}^a \pleft \Fermion{l}_{f_1,i_2} \antiscalar{l}_{f_1,i_1}
  \right) \tau^a_{i_1i_2}
  \\
  &\hphantom{{}=}
  -\frac{g_2^3}{4 \sqrt{2} \pi ^2}
  \left(
    \barFermion{{H_u}}_{i_1} \pright
    \Gaugino{W}^{a} \scalar{{H_u}}_{i_2} +
    \barGaugino{W}^a \pleft \Fermion{{H_u}}_{i_2} \antiscalar{{H_u}}_{i_1}
  \right) \tau^a_{i_1i_2}
  \\
  &\hphantom{{}=}
  -\frac{g_2^3}{4 \sqrt{2} \pi ^2}
  \left(
    \barFermion{{H_d}}_{i_1} \pright
    \Gaugino{W}^{a} \scalar{{H_d}}_{i_2} +
    \barGaugino{W}^a \pleft \Fermion{{H_d}}_{i_2} \antiscalar{{H_d}}_{i_1}
  \right) \tau^a_{i_1i_2}
  \\
  &\hphantom{{}=}
  -\frac{g_2^3}{4 \sqrt{2} \pi ^2}
  \left(
    \barFermion{Q}_{f_1,i_1,c_1} \pright
    \Gaugino{W}^{a} \scalar{Q}_{f_1,i_2,c_1} +
    \barGaugino{W}^a \pleft \Fermion{Q}_{f_1,i_2,c_1} \antiscalar{Q}_{f_1,i_1,c_1}
  \right) \tau^a_{i_1i_2}
  \\
  &\hphantom{{}=}
  -\frac{3 g_3^3}{8 \sqrt{2} \pi ^2}
  \left(
    \barFermion{Q}_{f_1,i_1,c_1} \pright
    \Gaugino{G}^{a} \scalar{Q}_{f_1,i_1,c_2} +
    \barGaugino{G}^a \pleft \Fermion{Q}_{f_1,i_1,c_2} \antiscalar{Q}_{f_1,i_1,c_1}
  \right) T^a_{c_1c_2}
  \\
  &\hphantom{{}=}
  + \frac{3 g_3^3}{8 \sqrt{2} \pi ^2}
  \left(
    \barGaugino{G}^a
    \pleft \CCFermion{u}_{f_1,c_2} \scalar{u}_{f_1,c_1} +
    \barCCFermion{u}_{f_1,c_1} \pright \Gaugino{G}^{a}
    \antiscalar{u}_{f_1,c_2}
  \right)T^a_{c_2c_1}
  \\
  &\hphantom{{}=}
  + \frac{3 g_3^3}{8 \sqrt{2} \pi ^2}
  \left(
    \barGaugino{W}^a
    \pleft \CCFermion{d}_{f_1,c_2} \scalar{d}_{f_1,c_1} +
    \barCCFermion{d}_{f_1,c_1} \pright \Gaugino{W}^{a}
    \antiscalar{d}_{f_1,c_2}
  \right)T^a_{c_2c_1}
\end{align*}
\begin{align*}
  \restLL{\text{gauge}} 
  &=
  -\frac{g_1 \left(g_1^2+27 g_2^2+48 g_3^2\right)}{3456 \pi ^2}
  \left(
    \barFermion{Q}_{f_1, i_1, c_1}
    \gamma_{\mu}
    \pleft \Fermion{Q}_{f_1, i_1, c_1}
    V_B^{\mu} 
  \right)
  \\
  &\hphantom{{}=} + 
  \frac{g_1^3+3 g_2^2 g_1}{128 \pi ^2}
  \left(\barFermion{l}_{f_1, i_1} 
    \gamma_{\mu}
  \pleft \Fermion{l}_{f_1, i_1}
  V_B^{\mu} \right)
\\
&\hphantom{{}=}
  -\frac{g_1 \left(g_1^2+3 g_2^2\right)}{128 \pi ^2}
  \left(\barFermion{{H_u}}_{i_1} 
  \gamma_{\mu}
  \pleft \Fermion{{H_u}}_{i_1}
  V_B^{\mu} \right)
  \\
  &\hphantom{{}=}
  + \frac{g_1 \left(g_1^2+3 g_2^2\right)}{128 \pi ^2}
  \left(\barFermion{{H_d}}_{i_1}
  \gamma_{\mu}
  \pleft \Fermion{{H_d}}_{i_1}
  V_B^{\mu} \right)
  \\
  &\hphantom{{}=}
  + \frac{g_1 \left(g_1^2+3 g_3^2\right)}{54 \pi ^2}
  \left(\barCCFermion{u}_{f_1,c_1}
  \gamma_{\mu}
  \pleft \CCFermion{u}_{f_1,c_1}
  V_B^{\mu} \right) 
  \\
  &\hphantom{{}=}
  -\frac{g_1 \left(g_1^2+12 g_3^2\right)}{432 \pi ^2}
  \left(\barCCFermion{d}_{f_1,c_1} 
  \gamma_{\mu}
  \pleft \CCFermion{d}_{f_1,c_1}
  V_B^{\mu} \right) 
  \\
  &\hphantom{{}=}
  -\frac{g_1^3}{16 \pi ^2}
  \left(\barCCFermion{e}_{f_1} 
  \gamma_{\mu}
  \pleft \CCFermion{e}_{f_1}
  V_B^{\mu} \right)
  \\
  &\hphantom{{}=}
  -\frac{g_2 \left(g_1^2+27 g_2^2+48 g_3^2\right)}{576 \pi ^2}
  \left(\barFermion{Q}_{f_1, i_1, c_1}
  \gamma^{\mu} 
  \pleft \Fermion{Q}_{f_1, i_2, c_1}
  \tau^a_{i_1i_2} \Vecbos{W}{a}{\mu} \right)
  \\
  &\hphantom{{}=}
  -\frac{g_2 \left(g_1^2+3 g_2^2\right)}{64 \pi ^2}
  \left(\barFermion{l}_{f_1, i_1}
  \gamma^{\mu} 
  \pleft \Fermion{l}_{f_1, i_2}
  \tau^a_{i_1i_2} \Vecbos{W}{a}{\mu}  \right)
  \\
  &\hphantom{{}=}
  -\frac{g_2 \left(g_1^2+3 g_2^2\right)}{64 \pi ^2}
  \left(\barFermion{{H_u}}_{i_1}
  \gamma^{\mu} 
  \pleft \Fermion{{H_u}}_{i2}
  \tau^a_{i_1i_2} \Vecbos{W}{a}{\mu} \right)
  \\
  &\hphantom{{}=}
  -\frac{g_2 \left(g_1^2+3 g_2^2\right)}{64 \pi ^2}
  \left(\barFermion{{H_d}}_{i_1}
  \gamma^{\mu} 
  \pleft \Fermion{{H_d}}_{i_2}
  \tau^a_{i_1i_2} \Vecbos{W}{a}{\mu} \right)
  \\
  &\hphantom{{}=}
  -\frac{g_3 \left(g_1^2+27 g_2^2+48 g_3^2\right)}{576 \pi ^2}
  \left(\barFermion{Q}_{f_1, i_1, c_1}
  \gamma^{\mu} 
  \pleft \Fermion{Q}_{f_1, i_1, c_2}
  T^a_{c_1c_2} \Vecbos{G}{a}{\mu} \right)
  \\
  &\hphantom{{}=}
  + \frac{g_3 \left(g_1^2+3 g_3^2\right)}{36 \pi ^2}
  \left(\barCCFermion{u}_{f_1,c_1} 
  \gamma^{\mu}
  \pleft \CCFermion{u}_{f_1,c_2}
  T^a_{c_2c_1}\Vecbos{G}{a}{\mu} \right) 
  \\
  &\hphantom{{}=}
  + \frac{g_3 \left(g_1^2+12 g_3^2\right)}{144 \pi ^2}
  \left(\barCCFermion{d}_{f_1,c_1} 
  \gamma^{\mu}
  \pleft \CCFermion{d}_{f_1,c_2}
  T^a_{c_2c_1}\Vecbos{G}{a}{\mu} \right) 
  \\
  &\hphantom{{}=}
  -\frac{\mathi g_2^3}{16 \pi ^2}
  \left(\barGaugino{W}^{a} 
  \gamma_{\mu} 
  \Gaugino{W}^{c}
  \varepsilon^{abc} \Vecbos{W}{b, \mu}{} \right)
  -\frac{3 \mathi g_3^3}{32 \pi ^2}
  \left(\barGaugino{G}^{a} 
  \gamma_{\mu} 
  \Gaugino{G}^{c}
  f^{abc} \Vecbos{G}{b, \mu}{} \right)
  \\
  &\hphantom{{}=}
  -\frac{g_2^3}{24 \pi ^2}
  \left(
    \varepsilon^{abc}
   \Vecbos{W}{a}{\mu} \Vecbos{W}{b}{\nu} 
  \left(\partial^{\nu}\Vecbos{W}{c,\mu}{}\right)
  \right)
  -\frac{g_3^3}{16 \pi ^2}
  \left(
  f^{abc}
  \Vecbos{G}{a}{\mu} \Vecbos{G}{b}{\nu} 
  \left(\partial^{\nu}\Vecbos{G}{c,\mu}{}\right)
  \right)
  \\
  &\hphantom{{}=}
  -\frac{g_2^4}{96 \pi ^2}
  \left(
  \varepsilon^{abc} \varepsilon^{ade}
  \Vecbos{W}{b}{\mu} \Vecbos{W}{d,\mu}{} \Vecbos{W}{c}{\nu} \Vecbos{W}{e,\nu}{}
  \right)
  \\
  &\hphantom{{}=}
  -\frac{g_3^4}{64 \pi ^2}
  \left(
  f^{abc} f^{ade}
  \Vecbos{G}{b}{\mu} \Vecbos{G}{d,\mu}{} \Vecbos{G}{c}{\nu} \Vecbos{G}{e,\nu}{}
  \right)
\end{align*}
\begin{align*}
  \restLL{\text{Yukawa}} 
  &=
  \frac{1}{32 \pi ^2}
  \Big\{
  \epsilon_{i_1i_2}^{\dagger} y^{e \dagger}_{f_2f_3}
  \Big[
    \left(g_1^2-3 g_2^2\right)
    \barFermion{l}_{f_2,i_2} \pright
    (\Fermion{{H_d}})^{\chargec}_{i_1} \scalar{e}_{f_3} 
    \\
    &\hphantom{{}=\frac{1}{32 \pi ^2} \Big\{ \epsilon_{i_1i_2}^{\dagger} y^{e \dagger}_{f_2f_3} \Big[ }
    - 2 g_1^2 
    \Big(
    \barCCFermion{e}_{f_3} \pright
    (\Fermion{l})^{\chargec}_{f_2,i_2} \antiscalar{{H_d}}_{i_1} 
    + \barFermion{{H_d}}_{i_1} \pright
    \Fermion{e}_{f_3} \antiscalar{l}_{f_2,i_2}
    \Big)
    \Big]
    \\
    &\hphantom{{}=\frac{1}{32 \pi ^2}\Big\{}
    + \epsilon_{i_1i_2} y^{e}_{f_2f_3} 
    \Big[
    \left(g_1^2-3 g_2^2\right)
    (\barFermion{{H_d}})^{\chargec}_{i_1} \pleft
    \Fermion{l}_{f_2,i_2} \antiscalar{e}_{f_3}
    \\
    &\hphantom{{}=\frac{1}{32 \pi ^2} \Big\{ + \epsilon_{i_1i_2} y^{e}_{f_2f_3} \Big[}
    - 2 g_1^2
    \Big(
    (\barFermion{l})^{\chargec}_{f_2,i_2} \pleft
    \CCFermion{e}_{f_3} \scalar{{H_d}}_{i_1} +
    \barFermion{e}_{f_3} \pleft
    \Fermion{{H_d}}_{i_1} \scalar{l}_{f_2,i_2}
    \Big)
    \Big]
    \Big\}
    \\
    &\hphantom{{}=}
    -\frac{1}{288 \pi ^2}
    \Big\{
    \epsilon_{i_1i_2}^{\dagger} y^{d \dagger}_{f_2f_3}
    \Big[
    3 \left(g_1^2+9 g_2^2\right)
    \barFermion{Q}_{f_2,i_2,c_2} \pright
    (\Fermion{{H_d}})^{\chargec}_{i_1} \scalar{d}_{f_3,c_2} 
    \\
    &\hphantom{{}=-\frac{1}{288 \pi ^2}\Big\{ \epsilon_{i_1i_2}^{\dagger} y^{d \dagger}_{f_2f_3} \Big[}
    -2 \left(g_1^2-24 g_3^2\right)
    \barCCFermion{d}_{f_3,c_2} \pright
    (\Fermion{Q})^{\chargec}_{f_2,i_2,c_2} \antiscalar{{H_d}}_{i_1}
    \\
    &\hphantom{{}=-\frac{1}{288 \pi ^2}\Big\{ \epsilon_{i_1i_2}^{\dagger} y^{d \dagger}_{f_2f_3} \Big[}
    + 6 g_1^2 
    \barFermion{{H_d}}_{i_1} \pright
    \Fermion{d}_{f_3,c_2} \antiscalar{Q}_{f_2,i_2,c_2}
    \Big]
    \\
    &\hphantom{{}=-\frac{1}{288 \pi ^2}\Big\{}
     + \epsilon_{i_1i_2} y^{d}_{f_2f_3} 
    \Big[
    3 \left(g_1^2+9 g_2^2\right)
    (\barFermion{{H_d}})^{\chargec}_{i_1} \pleft
    \Fermion{Q}_{f_2,i_2,c_2} \antiscalar{d}_{f_3,c_2}
    \\
    &\hphantom{{}=-\frac{1}{288 \pi ^2}\Big\{+ \epsilon_{i_1i_2} y^{d}_{f_2f_3} \Big[}
    -2 \left(g_1^2-24 g_3^2\right)
    (\barFermion{Q})^{\chargec}_{f_2,i_2,c_2} \pleft
    \CCFermion{d}_{f_3,c_2} \scalar{{H_d}}_{i_1}
    \\
    &\hphantom{{}=-\frac{1}{288 \pi ^2}\Big\{+ \epsilon_{i_1i_2} y^{d}_{f_2f_3} \Big[}
    + 6 g_1^2
    \barFermion{d}_{f_3,c_2} \pleft
    \Fermion{{H_d}}_{i_1} \scalar{Q}_{f_2,i_2,c_2}
    \Big]
    \Big\}
    \\
    &\hphantom{{}=}
    -\frac{1}{288 \pi ^2}
    \Big\{
    \epsilon^{\dagger}_{i_1i_2} y^{u \dagger}_{f_1f_3}
    \Big[
    4 \left(g_1^2+12 g_3^2\right)
    \barFermion{Q}_{f_1,i_1,c_3} \pright
    \Fermion{u}_{f_3,c_3} \antiscalar{{H_u}}_{i_2}
    \\
    &\hphantom{{}=-\frac{1}{288 \pi ^2}\Big\{ \epsilon^{\dagger}_{i_1i_2} y^{u \dagger}_{f_1f_3} \Big[}
    + 12 g_1^2
    \barCCFermion{u}_{f_3,c_3} \pright
    (\Fermion{{H_u}})^{\chargec}_{i_2} \antiscalar{Q}_{f_1,i_1,c_3}
    \\
    &\hphantom{{}=-\frac{1}{288 \pi ^2}\Big\{ \epsilon^{\dagger}_{i_1i_2} y^{u \dagger}_{f_1f_3} \Big[}
    + 3 \left( g_1^2 - 9 g_2^2 \right)
    \barFermion{{H_u}}_{i_2} \pright
    (\Fermion{Q})^{\chargec}_{f_1,i_1,c_3} 
    \scalar{u}_{f_3,c_3}
    \Big]
    \\
    &\hphantom{{}=-\frac{1}{288 \pi ^2}\Big\{}
    + \epsilon_{i_1i_2} y^{u}_{f_1f_3}
    \Big[
    + 4 \left(g_1^2+12 g_3^2\right)
    \barFermion{u}_{f_3,c_3} \pleft
    \Fermion{Q}_{f_1,i_1,c_3} \scalar{{H_u}}_{i_2}
    \\
    &\hphantom{{}=-\frac{1}{288 \pi ^2}\Big\{+ \epsilon_{i_1i_2} y^{u}_{f_1f_3} \Big[}
    + 12 g_1^2
    (\barFermion{{H_u}})^{\chargec}_{i_2} \pleft
    \CCFermion{u}_{f_3,c_3} \scalar{Q}_{f_1,i_1,c_3}
    \\
    &\hphantom{{}=-\frac{1}{288 \pi ^2}\Big\{+ \epsilon_{i_1i_2} y^{u}_{f_1f_3} \Big[}
    + 3 \left( g_1^2 - 9 g_2^2 \right)
    (\barFermion{Q})^{\chargec}_{f_1,i_1,c_3} \pleft
    \Fermion{{H_u}}_{i_2} \antiscalar{u}_{f_3,c_3}
    \Big]
    \Big\}
\end{align*}
\begin{align*}
  \restLL{\text{quartic}} 
  &=
  \frac{1}{5184 \pi ^2}
  \scalar{d}_{f_1,c_1} \scalar{d}_{f_2,c_2}
  \Big[
    9 g_3^2 \left(8 g_1^2+15 g_3^2\right)
    \antiscalar{d}_{f_1,c_2} \antiscalar{d}_{f_2,c_1}
    \\
    &\hphantom{{}= \frac{1}{5184 \pi ^2} \scalar{d}_{f_1,c_1} \scalar{d}_{f_2,c_2} \Big[}
    + \left(8 g_1^4-24 g_3^2 g_1^2+99 g_3^4\right)
    \antiscalar{d}_{f_1,c_1} \antiscalar{d}_{f_2,c_2}
  \Big]
  \\
  &\hphantom{{}=}
  + \frac{g_1^4}{144 \pi^2} 
  \scalar{d}_{f_1,c_1} \antiscalar{d}_{f_1,c_1}
  \scalar{{H_d}}_{i_2} \antiscalar{{H_d}}_{i_2}
  + \frac{g_1^4}{144 \pi^2} 
  \scalar{d}_{f_1,c_1} \antiscalar{d}_{f_1,c_1}
  \scalar{{H_u}}_{i_2} \antiscalar{{H_u}}_{i_2}
  \\
  &\hphantom{{}=}
  + \frac{g_1^4}{144 \pi^2} 
  \scalar{d}_{f_1,c_1} \antiscalar{d}_{f_1,c_1}
  \scalar{l}_{f_2,i_2} \antiscalar{l}_{f_2,i_2}
  \\
  &\hphantom{{}=}
  + \frac{1}{2592 \pi ^2}\scalar{d}_{f_1,c_1} \scalar{Q}_{f_2,i_2,c_2}
  \Big[
  9 g_3^2 \left(15 g_3^2-4 g_1^2\right) 
  \antiscalar{d}_{f_1,c_2} \antiscalar{Q}_{f_2,i_2,c_1}
  \\
  &\hphantom{{}=+ \frac{1}{2592 \pi ^2}\scalar{d}_{f_1,c_1} \scalar{Q}_{f_2,i_2,c_2} \Big[}
  + \left(2 g_1^4+12 g_3^2 g_1^2+99 g_3^4\right)
  \antiscalar{d}_{f_1,c_1} \antiscalar{Q}_{f_2,i_2,c_2}
  \Big]
  \\
  &\hphantom{{}=}
  + \frac{1}{2592 \pi ^2} \scalar{d}_{f_1,c_1} \scalar{u}_{f_2,c_2}
  \Big[
  9 g_3^2 \left(15 g_3^2-16 g_1^2\right)
  \antiscalar{d}_{f_1,c_2} \antiscalar{u}_{f_2,c_1}
  \\
  &\hphantom{{}=+ \frac{1}{2592 \pi ^2} \scalar{d}_{f_1,c_1} \scalar{u}_{f_2,c_2}\Big[}
  + \left(32 g_1^4+48 g_3^2 g_1^2+99 g_3^4\right)
  \antiscalar{d}_{f_1,c_1} \antiscalar{u}_{f_2,c_2}
  \Big]
  \\
  &\hphantom{{}=}
  + \frac{g_1^4}{36 \pi^2} 
  \scalar{d}_{f_2,c_2} \antiscalar{d}_{f_2,c_2} 
  \scalar{e}_{f_1} \antiscalar{e}_{f_1}  
  + \frac{g_1^4}{8 \pi^2}
  \scalar{e}_{f_1} \antiscalar{e}_{f_1} 
  \scalar{e}_{f_2} \antiscalar{e}_{f_2}
  \\
  &\hphantom{{}=}
  + \frac{g_1^4}{16 \pi^2}
  \scalar{e}_{f_1} \antiscalar{e}_{f_1} 
  \scalar{{H_d}}_{i_2} \antiscalar{{H_d}}_{i_2}
  + \frac{g_1^4}{16 \pi^2}
  \scalar{e}_{f_1} \antiscalar{e}_{f_1} 
  \scalar{{H_u}}_{i_2} \antiscalar{{H_u}}_{i_2}
  \\
  &\hphantom{{}=}
  + \frac{g_1^4}{16 \pi^2}
  \scalar{e}_{f_1} \antiscalar{e}_{f_1} 
  \scalar{l}_{f_2,i_2} \antiscalar{l}_{f_2,i_2}
  + \frac{g_1^4}{144 \pi^2}
  \scalar{e}_{f_1} \antiscalar{e}_{f_1} 
  \scalar{Q}_{f_2,i_2,c_2} \antiscalar{Q}_{f_2,i_2,c_2}
  \\
  &\hphantom{{}=}
  + \frac{g_1^4}{9 \pi^2}
  \scalar{e}_{f_1} \antiscalar{e}_{f_1} 
  \scalar{u}_{f_2,c_2} \antiscalar{u}_{f_2,c_2}
  \\
  &\hphantom{{}=}
  + \frac{1}{128 \pi^2} 
  \scalar{{H_d}}_{i_3} \antiscalar{{H_d}}_{i_1} 
  \scalar{{H_d}}_{i_4} \antiscalar{{H_d}}_{i_2}
  \\
  &\hphantom{{}=+} \times
  \Big[
   g_1^4 \delta_{i_1i_4} \delta_{i_2i_3} 
  + 4 g_2^2
  \Big(
  2 g_1^2\bigl(\tau^a_{i_1i_4} \tau^a_{i_2i_3} \bigr)
  +  g_2^2\bigl(\{\tau^a,\tau^b\}_{i_1i_3}\{\tau^a,\tau^b\}_{i_2i_4} 
  \bigr) \Big) \Big]
  \\
  &\hphantom{{}=}
  + \frac{1}{64 \pi^2}
  \scalar{{H_d}}_{i_3} \antiscalar{{H_d}}_{i_1} 
  \scalar{{H_u}}_{i_4} \antiscalar{{H_u}}_{i_2}
  \\
  &\hphantom{{}=+ } \times
  \Big[
  g_1^4 \delta_{i_1i_3} \delta_{i_2i_4} 
  + 4 g_2^2
  \Big(
  - 2 g_1^2\bigl( 
    \tau^a_{i_1i_3} \tau^a_{i_2i_4} \bigr)
  + g_2^2\bigl(\{\tau^a,\tau^b\}_{i_1i_3}\{\tau^a,\tau^b\}_{i_2i_4} 
  \bigr) \Big) \Big]
\\
  &\hphantom{{}=}
  + \frac{1}{64 \pi^2}
  \scalar{{H_d}}_{i_3} \antiscalar{{H_d}}_{i_1} 
  \scalar{l}_{f_2,i_4} \antiscalar{l}_{f_2,i_2}
  \\
  &\hphantom{{}=+ } \times
  \Big[
  g_1^4 \delta_{i_1i_3} \delta_{i_2i_4} 
  + 4 g_2^2
  \Big(
  2 g_1^2\bigl( 
    \tau^a_{i_1i_3} \tau^a_{i_2i_4} \bigr)
  + g_2^2\bigl(\{\tau^a,\tau^b\}_{i_1i_3}\{\tau^a,\tau^b\}_{i_2i_4} 
  \bigr) \Big) \Big]
  \\
  &\hphantom{{}=}
  + \frac{1}{576 \pi^2}
  \scalar{{H_d}}_{i_3} \antiscalar{{H_d}}_{i_1} 
  \scalar{Q}_{f_2,i_4,c_2} \antiscalar{Q}_{f_2,i_2,c_2}
  \\
  &\hphantom{{}=+ } \times
  \Big[
  g_1^4 \delta_{i_1i_3} \delta_{i_2i_4} 
  + 12 g_2^2
  \Big(
  - 2 g_1^2\bigl( 
    \tau^a_{i_1i_3} \tau^a_{i_2i_4} \bigr)
  + 3 g_2^2\bigl(\{\tau^a,\tau^b\}_{i_1i_3}\{\tau^a,\tau^b\}_{i_2i_4} 
  \bigr) \Big) \Big]
  \\
  &\hphantom{{}=}
  + \frac{1}{128 \pi^2} 
  \scalar{H_u}_{i_3} \antiscalar{{H_u}}_{i_1} 
  \scalar{{H_u}}_{i_4} \antiscalar{{H_u}}_{i_2}
  \\
  &\hphantom{{}=+ } \times
  \Big[
   g_1^4 \delta_{i_1i_4} \delta_{i_2i_3} 
  + 4 g_2^2
  \Big(
  2 g_1^2\bigl( \tau^a_{i_1i_4} \tau^a_{i_2i_3} \bigr)
  +  g_2^2\bigr(\{\tau^a,\tau^b\}_{i_1i_3}\{\tau^a,\tau^b\}_{i_2i_4} 
  \bigl) \Big) \Big]
  \\
  &\hphantom{{}=}
  + \frac{1}{64 \pi^2}
  \scalar{{H_u}}_{i_3} \antiscalar{{H_u}}_{i_1} 
  \scalar{l}_{f_2,i_4} \antiscalar{l}_{f_2,i_2}
  \\
  &\hphantom{{}=+ } \times
  \Big[
  g_1^4 \delta_{i_1i_3} \delta_{i_2i_4} 
  + 4 g_2^2
  \Big(
  - 2 g_1^2\bigl( 
    \tau^a_{i_1i_3} \tau^a_{i_2i_4} \bigr)
  + g_2^2\bigl(\{\tau^a,\tau^b\}_{i_1i_3}\{\tau^a,\tau^b\}_{i_2i_4} 
  \bigr) \Big) \Big]
  \\
  &\hphantom{{}=}
  + \frac{1}{576 \pi^2}
  \scalar{{H_u}}_{i_3} \antiscalar{{H_u}}_{i_1} 
  \scalar{Q}_{f_2,i_4,c_2} \antiscalar{Q}_{f_2,i_2,c_2}
  \\
  &\hphantom{{}=+ } \times
  \Big[
  g_1^4 \delta_{i_1i_3} \delta_{i_2i_4} 
  + 12 g_2^2
  \Big(
  2 g_1^2\bigl( 
    \tau^a_{i_1i_3} \tau^a_{i_2i_4} \bigr)
  + 3 g_2^2\bigr(\{\tau^a,\tau^b\}_{i_1i_3}\{\tau^a,\tau^b\}_{i_2i_4} 
  \bigr) \Big) \Big]
  \\
  &\hphantom{{}=}
  + \frac{1}{128 \pi^2}
  \scalar{l}_{f_1,i_3} \antiscalar{l}_{f_1,i_1} 
  \scalar{l}_{f_2,i_4} \antiscalar{l}_{f_2,i_2}
  \\
  &\hphantom{{}=+ } \times
  \Big[
  g_1^4 \delta_{i_1i_3} \delta_{i_2i_4} 
  + 4 g_2^2
  \Big(
  g_1^2 \bigl(\tau^a_{i_1i_3} \tau^a_{i_2i_4}\bigr)
  +  g_2^2 \bigl(\{\tau^a,\tau^b\}_{i_1i_3} 
  \{\tau^a,\tau^b\}_{i_2i_4} \bigr)
  \Big) \Big]
  \\
  &\hphantom{{}=}
  + \frac{1}{576 \pi^2}
  \scalar{l}_{f_1,i_3} \antiscalar{l}_{f_1,i_1} 
  \scalar{Q}_{f_2,i_4,c_2} \antiscalar{Q}_{f_2,i_2,c_2}
  \\
  &\hphantom{{}=+ } \times
  \Big[
  g_1^4 \delta_{i_1i_3} \delta_{i_2i_4} 
  + 12 g_2^2
  \Big(
  - 2 g_1^2\bigl( 
    \tau^a_{i_1i_3} \tau^a_{i_2i_4} \bigr)
  + 3 g_2^2\bigl(\{\tau^a,\tau^b\}_{i_1i_3}\{\tau^a,\tau^b\}_{i_2i_4} 
  \bigr) \Big) \Big]
  \\
  &\hphantom{{}=}
  + \frac{g_1^4}{36 \pi^2} 
  \scalar{{H_d}}_{i_2} \antiscalar{{H_d}}_{i_2} 
  \scalar{u}_{f_1,c_1} \antiscalar{u}_{f_1,c_1}
  + \frac{g_1^4}{36 \pi^2} 
  \scalar{{H_u}}_{i_2} \antiscalar{{H_u}}_{i_2} 
  \scalar{u}_{f_1,c_1} \antiscalar{u}_{f_1,c_1}
  \\
  &\hphantom{{}=}
  + \frac{g^4_1}{36 \pi^2}
  \scalar{l}_{f_2,i_2} \antiscalar{l}_{f_2,i_2} 
  \scalar{u}_{f_1,c_1} \antiscalar{u}_{f_1,c_1}
\\
  &\hphantom{{}=}
  + \frac{1}{2592 \pi ^2}\scalar{u}_{f_1,c_1} \scalar{Q}_{f_2,i_2,c_2}
  \Big[
  \left(8 g_1^4-24 g_3^2 g_1^2+99 g_3^4\right)
  \antiscalar{u}_{f_1,c_1} \antiscalar{Q}_{f_2,i_2,c_2}
  \\
  &\hphantom{{}=+ \frac{1}{2592 \pi ^2}\scalar{u}_{f_1,c_1} \scalar{Q}_{f_2,i_2,c_2} \Big[}
  + 9 g_3^2 \left(8 g_1^2+15 g_3^2\right)
  \antiscalar{u}_{f_1,c_2} \antiscalar{Q}_{f_2,i_2,c_1}
  \Big]
  \\
  &\hphantom{{}=}
  + \frac{1}{5184 \pi ^2}
  \scalar{u}_{f_1,c_1} \scalar{u}_{f_2,c_2}
  \Big[
    9 g_3^2 \left(32 g_1^2+15 g_3^2\right)
    \antiscalar{u}_{f_1,c_2} \antiscalar{u}_{f_2,c_1}
    \\
    &\hphantom{{}= + \frac{1}{5184 \pi ^2} \scalar{u}_{f_1,c_1} \scalar{u}_{f_2,c_2} \Big[}
    + \left(128 g_1^4-96 g_3^2 g_1^2+99 g_3^4\right)
    \antiscalar{u}_{f_1,c_1} \antiscalar{u}_{f_2,c_2}
  \Big]
  \\
  &\hphantom{{}=}
  + \frac{1}{10368 \pi^2} 
  \scalar{Q}_{f_1,i_3,c_3} \antiscalar{Q}_{f_1,i_1,c_1} 
  \scalar{Q}_{f_2,i_4,c_4} \antiscalar{Q}_{f_2,i_2,c_2}
  \\
  &\hphantom{{}=+ }
  \times \Big[
  18 g_3^2 \left( 2 g_1^2 + 15 g_3^2 \right)
  \delta_{i_1i_3} \delta_{i_2i_4} \delta_{c_2c_3} \delta_{c_1c_4}
  \\
  &\hphantom{{}=+ \times\Big\{}
  + \left(g_1^4 - 12 g_1^2 g_3^2 + 198 g_3^4\right)
  \delta_{i_1i_3} \delta_{i_2i_4} \delta_{c_1c_3} \delta_{c_2c_4}
  \\
  &\hphantom{{}=+ \times\Big\{}
  + 18 g_2^2 
  \Big(
  48 g_3^2 \delta_{c_2c_3} \delta_{c_1c_4}
  \tau^a_{i_1i_3} \tau^a_{i_2i_4}
  + 4  g_1^2 \delta_{c_1c_3} \delta_{c_2c_4}
  \tau^a_{i_1i_3} \tau^a_{i_2i_4}
  \\
  &\hphantom{{}=+ \times\Big\{+ 18 g_2^2 \Big[}
  + 9 g_2^2 \delta_{c_1c_3} \delta_{c_2c_4}
  \bigl(\{\tau^a,\tau^b\}_{i_1i_3}\{\tau^a,\tau^b\}_{i_2i_4}
    + 2 \left(\tau^a \tau^b\right)_{i_2i_4} \{\tau^a,\tau^b\}_{i_1i_3}
  \bigr)
  \Big)
  \Big]
\end{align*}

\end{appendix}

\bibliographystyle{elsarticle-num}
\bibliography{<your-bib-database>}

\end{document}